\documentclass[letterpaper, 10 pt, conference]{ieeeconf}
\IEEEoverridecommandlockouts                              % This command is only
\usepackage{ntheorem}
\usepackage{amsmath}
\usepackage{amsfonts}
\usepackage{listings}
\usepackage{color}
\usepackage{array}
\usepackage{tabu}
\usepackage{mathrsfs,amssymb}
\usepackage{graphicx}
\graphicspath{ {image22/} }
\usepackage{float}
\usepackage{algorithm}

{}
{}
{}
\newtheorem{theorem}{Theorem}{}
\newtheorem{remark}{Remark}{}
\newtheorem{lemma}{Lemma}{}
\newtheorem{assumption}{Assumption}{}
{}

\usepackage{algpseudocode}
\algdef{SE}[DOWHILE]{Do}{doWhile}{\algorithmicdo}[1]{\algorithmicwhile\ #1}%
\DeclareMathOperator*{\argmin}{arg\,min}

\DeclareMathOperator*{\card}{card}
\DeclareMathOperator*{\supp}{supp}

\DeclareMathOperator*{\rank}{rank}
\title{\LARGE \bf
An Unknown Input Multi-Observer Approach for Estimation, Attack Isolation, and Control of LTI Systems under Actuator Attacks
}

\author{Tianci Yang, Carlos Murguia, Margreta Kuijper, and Dragan Ne\v{s}i\'{c} % <-this % stops a space
\thanks{This work was partially supported by the Australian Research Council under the Discovery Project DP170104099. Tianci Yang, Carlos Murguia, Margreta Kuijper, and Dragan Ne\v{s}i\'{c} are with the Department of Electrical and Electronic Engineering, at the University of Melbourne, Australia. Emails: tianciy@student.unimelb.edu.au, \ carlos.murguia@unimelb.edu.au, \ mkuijper@unimelb.edu.au,  \& dnesic@unimelb.edu.au.}
}

\begin{document}
\maketitle
\thispagestyle{empty}
\pagestyle{empty}

\begin{abstract}
We address the problem of state estimation, attack isolation, and control for discrete-time Linear Time Invariant (LTI) systems under (potentially unbounded) actuator false data injection attacks. Using a bank of Unknown Input Observers (UIOs), each observer leading to an exponentially stable estimation error in the attack-free case, we propose an estimator that provides exponential estimates of the system state and the attack signals when a sufficiently small number of actuators are attacked. We use these estimates to control the system and isolate actuator attacks. Simulations results are presented to illustrate the performance of the results.
\end{abstract}


% \fi
\IEEEpeerreviewmaketitle

\section{Introduction}
Networked Control Systems (NCSs) have emerged as a technology that combines control, communication, and computation and offers the necessary flexibility to meet new demands in distributed and large scale systems. Recently, security of NCSs has become an important issue as wireless communication networks might serve as new access points for attackers to adversely affect the operation of the system dynamics. Cyber-physical attacks on NCSs have caused substantial damage to a number of physical processes. One of the most well-known examples is the attack on Maroochy Shire Council’s sewage control system in Queensland, Australia that happened in January 2000. The attacker hacked into the controllers that activate and deactivate valves and caused flooding of the grounds of a hotel, a park, and a river with a million liters of sewage. Another incident is the more recent StuxNet virus that targeted Siemens’ supervisory control and data acquisition systems which are used in many industrial processes. It follows that strategic mechanisms to identify and deal with attacks on NCSs are strongly needed.

In \cite{Fawzi2012}-\nocite{Massoumnia1986}\nocite{Pajic2014a}\nocite{Mo2014}\nocite{Vamvoudakis2014}\nocite{Chong2016b}\nocite{Vamvoudakis2012}\nocite{Shoukry2014}\nocite{Yong}\nocite{Park2015}\nocite{Liu2009}\nocite{Kuijper2018}\nocite{Teixeira2012b}\nocite{Murguia2016}\nocite{Dolk1}\nocite{Hashemil2017}\nocite{Pasqualetti123}\nocite{Murguia2017d}\nocite{Jairo}\nocite{Carlos_Justin2}\nocite{Sahand2017}\nocite{Carlos_Justin3}\cite{Carlos_Iman1}, a range of topics related to security of linear control systems have been discussed. In general, they provide analysis tools for quantifying the performance degradation induced by different classes of attacks; and propose reaction strategies to identify and counter their effect on the system dynamics. There are also some results addressing the nonlinear case. In \cite{Kim2016a}, exploiting sensor redundancy, the authors address the problem of sensor attack detection and state estimation for uniformly observable continuous-time nonlinear systems. Similarly, in \cite{tianci1}, the authors provide an algorithm for isolating sensor attacks for a class of discrete-time nonlinear systems with bounded measurement noise.

In this manuscript, we use Unknown Input Observers (UIOs) to address the problem of state estimation, attack isolation, and control for discrete-time Linear Time Invariant (LTI) systems under (potentially unbounded) actuator attacks. Unknown input observers are dynamical systems capable of estimating the state of the plant \emph{without} using input signals. If such an observer exists and some of the inputs are subject to attacks, we can reconstruct the system state without using inputs; and use these estimates to reconstruct the attack signals by using model matching techniques. The existence of UIOs depend on the system dynamics, i.e., the matrices $(A,B,C)$ comprising the system. If an UIO does not exists for this \emph{complete} $(A,B,C)$ but it does for some \emph{partial} $(A,\tilde{B}_i,C)$, where $\tilde{B}_i$ denotes a submatrix of $B$ with fewer columns and the same number of rows, then, using \emph{a bank of observers}, we can use similar ideas to perform state estimation and attack isolation at the price of only being able to isolate when a sufficiently small subset of actuators are under attack. The main idea behind our multi-observer estimator is the following. Each UIO in the bank is constructed using a triple $(A,\tilde{B}_i,C)$, i.e., the $i$-th observer does \emph{not} use the input signals associated with $\tilde{B}_i$, but it does use the remaining input signals. If the inputs corresponding to $\tilde{B}_i$ include all the attacked ones, this UIO produces an exponentially stable estimation error. For every pair of UIOs in the bank, we compute the largest difference between their estimates. Then, we select the pair leading to the smallest difference and prove that these observers reconstruct the state of the system exponentially. This idea is guaranteed to work under the assumption that less than half of the actuators are attacked. This multi-observer approach is inspired by the results given in \cite{Chong2015c} where the problem of state estimation for continuous-time LTI systems under sensor attacks is considered. Once we have an estimate of the state, we reconstruct the attack signals using model matching techniques. Finally, we propose a simple yet effective technique to stabilize the system by switching off the isolated actuators, and closing the loop with a multi-observer based output dynamic feedback controller. Here, we assume that the set of attacked actuators is time-invariant, i.e., if the opponent compromises a set of actuators at some time-instant, only this set will be compromised in forward time. Because attack signals may be zero for some time instants, the actuators isolated as attack-free might arbitrarily switch among all the supersets of the set of attack-free actuators. Therefore, we need a controller able to stabilize the closed-loop dynamics under the arbitrary switching induced by turning off the isolated actuators. To achieve this, we assume that a \textit{state} feedback controller that stabilizes the switching closed-loop system exists, and use this controller together with the multi-observer estimator to stabilize the system. We use Input-to-State Stability (ISS) \cite{Sontag2008} of the closed-loop system with respect to the exponentially stable estimation error to conclude on stability of the closed-loop dynamics.

The paper is organized as follows. In Section \uppercase\expandafter{\romannumeral2}, we present some preliminary results needed for the subsequent sections. In Section \uppercase\expandafter{\romannumeral3}, we introduce the proposed UIO-based estimation schemes. In Section \uppercase\expandafter{\romannumeral4}, a method for isolating actuator attacks is described. The proposed control scheme is given in Section \ref{control}. Finally, in Section \uppercase\expandafter{\romannumeral6}, we give concluding remarks.

\section{Preliminaries}
\subsection{Notation}
We denote the set of real numbers by $\mathbb{R}$, the set of natural numbers by $\mathbb{N}$ , the set of integers by $\mathbb{Z}$, and the set of $n\times m$ real matrices by $\mathbb{R}^{n\times m}$, $m,n \in \mathbb{N}$. For any vector $v\in\mathbb{R}^{n_{v}}$, {$v_{J}$} denotes the stacking of all $v_{i}$, $i\in J$ and $J\subset \left\lbrace 1,\ldots,n_{v}\right\rbrace$, $|v|=\sqrt{v^{\top} v}$ and $\supp(v)=\left\lbrace i\in\left\lbrace 1,\ldots,n_{v}\right\rbrace |v_{i}\neq0\right\rbrace $. For a sequence of vectors $\left\lbrace v(k)\right\rbrace _{k=0}^{\infty}$, we denote by $v_{[0,k]}$ the sequence of vectors $v(i)$, $i=0,\ldots,k$,  $||v||_{\infty} := \sup_{k\geq 0}|v(k)|$ and $||v||_{T} := \sup_{0\leq k\leq T}|v(k)|$. We say that a sequence $\left\lbrace v(k)\right\rbrace \in l_{\infty}$, if $||v||_{\infty}<\infty$. We denote the cardinality of a set $S$ as $\card(S)$. The binomial coefficient is denoted as $\binom{a}{b}$, where $a,b$ are nonnegative integers. We denote a variable $m$ uniformly distributed in the interval $(z_{1},z_{2})$ as $m\sim\mathcal{U}(z_{1},z_{2})$ and normally distributed with mean $\mu$ and variance $\sigma^2$ as $m\sim \mathcal{N}(\mu,\sigma^2)$. The notation $\mathbf{0}_{n}$ and $I_{n}$ denote the zero matrix and the identity matrix of dimension $\mathbb{R}^{n \times n}$, respectively. We simply write $\mathbf{0}$ and $I$ when their dimensions are evident.

\section{Unknown input observer-based estimator}
Consider the discrete-time LTI system:
\begin{eqnarray}\label{s1}
\left\{ \begin{split}
x^{+}=&Ax+B(u+a),\\
y=&Cx,
\end{split} \right.
\end{eqnarray}
with state $x\in\mathbb{R}^{n}$, output $y\in\mathbb{R}^{n_{y}}$, known input $u\in\mathbb{R}^{p}$, and vector of actuator attacks $a \in \mathbb{R}^{p}$, $a = (a_1,\ldots,a_p)^T$. That is, $a_{i}(k)=0$ for all $k\geq 0$ if the $i$-th actuator is attack-free; otherwise, $a_{i}(k) \neq 0$ for some (but not necessarily all) time instants $k\geq 0$, and can be arbitrarily large. Matrices $A,B,C$ are of appropriate dimensions, $(A,B)$ is stabilizable, $(A,C)$ is detectable, and $B$ has full column rank. Let the set of \emph{unknown} attacked actuators be denoted by $W\subset\left\lbrace 1,\hdots,p\right\rbrace$, i.e., $a_{i}(k_{i})\neq 0$ for some $k_{i}\geq 0$ and all $i\in W$.

\begin{assumption}\label{asump2}
	The set of attacked actuators is time invariant, i.e., $W \subset\left\lbrace 1,\hdots,p\right\rbrace $ is a constant set.
%	if the set of attack actuators at time $k=k^*$ is $W(k^*)$; then, $W(k) = W(k^*)$ for all $k \geq k^*$.
\end{assumption}

\subsection{Complete Unknown Input Observers}\label{complete}
\begin{figure}[t]
	\includegraphics[width=0.5\textwidth]{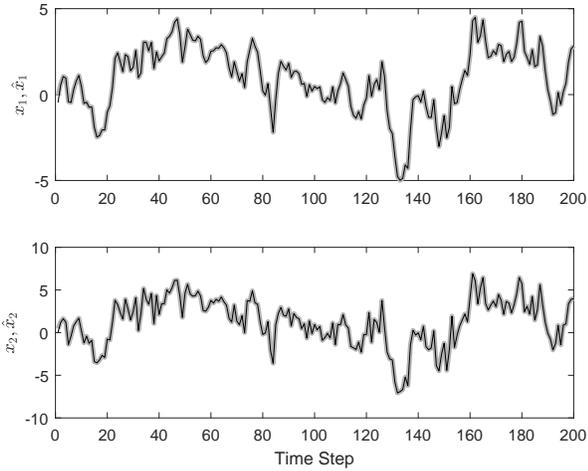}
	\caption{Estimated states $\hat{x}$ converges to the true states $x$ when $a\sim\mathcal{U}(-1,1)$. Legend: $\hat{x}$ (grey), true states (black)}
	\label{fig:1}
	\centering
\end{figure}

We first treat $(u+a)$ as an unknown input to system (\ref{s1}) and consider an UIO with the following structure:
\begin{eqnarray}\label{o}
\left\{ \begin{split}
z^{+}=&Nz+Ly,\\
\hat{x}=&z+Ey,
\end{split} \right.
\end{eqnarray}
where $z \in\mathbb{R}^{n}$ is the state of the observer, $\hat{x} \in \mathbb{R}^{n}$ denotes the estimate of the system state, and $(N,L,E)$ are observer matrices of appropriate dimensions to be designed. It is easy to verify that if $(N,L,E)$ satisfy the following equations:
\begin{eqnarray}\label{eq}
%\left\{ \begin{split}
%N&=A-ECA-L_{1}C,\\
%%T&=I-EC\\
%L&=NE,\\
%(EC-I)B&=\mathbf{0},\\
%L&=L_{1}+L_{2};
	\left\{ \begin{split}
	N(I-EC)+LC+(EC-I)A=&\mathbf{0},\\
	(EC-I)B=&\mathbf{0},
\end{split} \right.
\end{eqnarray}
then, the estimation error $e :=\hat{x}-x$ satisfies the difference equation:
\begin{equation}\label{error_Dyn1}
e^{+}=Ne.
\end{equation}
Hence, if $N$ is Schur, system (\ref{o}) is an UIO for (\ref{s1}). In \cite{Ding2008}, it is proved that such an observer exists if and only if the following two conditions are satisfied:\\[1mm]
\textbf{($\text{c}_\text{1}$)} $\rank(CB)=\rank(B)=p$.\\[1mm]
\textbf{($\text{c}_\text{2}$)} Matrix $E$ in (\ref{o}) yields the pair $(A-ECA,C)$ detectable.\\[1mm]
Assume that conditions ($\text{c}_\text{1}$) and ($\text{c}_\text{2}$) are satisfied; then, observer (\ref{o}) can be constructed by solving (\ref{eq}) for a Schur matrix $N$. Hence, for such an observer, there exist $c>0$ and $\lambda\in(0,1)$ satisfying:
\begin{equation}
|\hat{x}(k)-x(k)|\leq c\lambda^{k}|\hat{x}(0)-x(0)|,
\end{equation}
for $k\geq 0$, i.e., observer (\ref{o}) reconstructs the system state without using any input for arbitrarily large attack signals $a$.\\[2mm]
\textbf{Example 1:} Consider the following system:
\begin{eqnarray}
\left\{
\begin{split}
x^{+}=&\left[ \begin{matrix}\label{e1}
0.2&0.5\\
0.2&0.7\\
\end{matrix}\right] x+\left[ \begin{matrix}
1\\
2\\
\end{matrix}\right] (u+a),\\
y=&\left[ \begin{matrix}
1&3\\1&1\\3&2
\end{matrix}\right] x.
\end{split}
\right.
\end{eqnarray}
%An UIO exists for system (\ref{e1}). We let $W=\left\lbrace 1,2,3\right\rbrace $, i.e., all actuators are attacked, $a_{1},a_{2},a_{3} \in \mathcal{U}(-1,1)$, $u_{1},u_{2},u_{3} \in \mathcal{U}(-1,1)$, and initial conditions $x_{1}(0),x_{2}(0),x_{3}(0)\in\mathcal{N}(0,1)$. We solve \eqref{eq} for some Schur matrix $N$ and construct an unknown input observer for (\ref{e1}). The performance of the estimator$  $ is shown in Figure \ref{fig:1} for $\hat{x}(0)=\left[ 0,0,0\right] ^{\top}$.
An UIO exists for system (\ref{e1}). We let $W=\left\lbrace 1\right\rbrace $, $a\in \mathcal{U}(-1,1)$, $u\in \mathcal{U}(-1,1)$, and the initial conditions $x_{1}(0),x_{2}(0)\in\mathcal{N}(0,1)$. We solve \eqref{eq} for some Schur matrix $N$ and construct an unknown input observer for (\ref{e1}). The performance of the estimator is shown in Figure \ref{fig:1} for $\hat{x}(0)=\left[ 0,0\right] ^{\top}$.

\subsection{Partial Unknown Input Observers}\label{partial}
\begin{figure}[t]
	\includegraphics[width=0.5\textwidth]{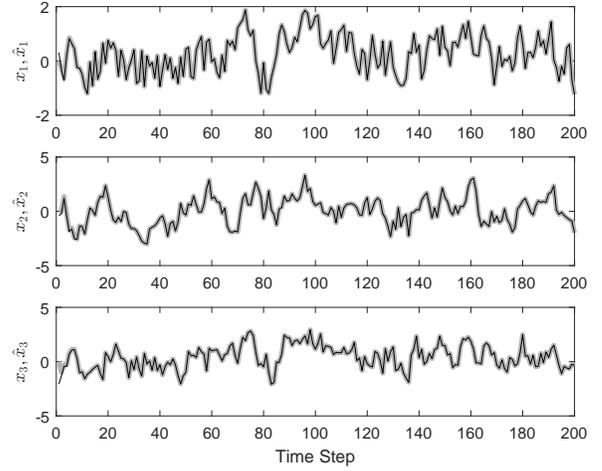}
	\caption{Estimated states $\hat{x}$ converges to the true states $x$ when $a_{2}\sim\mathcal{U}(-1,1)$. Legend: $\hat{x}$ (grey), true states (black)}
	\label{fig:1e}
	\centering
\end{figure}

In \cite{Chong2015c}, the problem of state estimation for continuous-time LTI system under sensor attacks is solved using a bank of Luenberger observers. Inspired by these results, we use a bank of partial UIOs to estimate the state of the system when actuator attacks occur.
Here, we are implicitly assuming that either condition ($\text{c}_\text{1}$) or ($\text{c}_\text{2}$) (or both) cannot be satisfied by the triple $(A,B,C)$. Let $B$ be partitioned as $B=\left[ b_{1},\hdots,b_{i},\hdots,b_{p}\right] $ where
$b_{i}\in\mathbb{R}^{n\times 1}$ is the $i$-th column of $B$. Then, the attacked system (\ref{s1}) can be written as
\begin{eqnarray}\label{s2}
\left\{
\begin{split}
x^{+}=&Ax+Bu+b_{W}a_{W},  \\
y=&Cx,
\end{split}
\right.
\end{eqnarray}
where the attack input $a_{W}$ can be regarded as an unknown input and the columns of $b_{W}$ are $b_{i}$ for $i\in W$. Denote by $b_{J}$ the matrix whose columns are $b_{i}$ for $i \in J$. Let $q$ be the largest integer such that for all $J\subset\left\lbrace 1,\ldots,p\right\rbrace $ with $\card(J)\leq 2q$, the following is satisfied:\\[1mm]
\textbf{($\text{c}_\text{3}$)} $\rank(Cb_{J})=\rank(b_{J})=\card(J)$.\\[1mm]
\textbf{($\text{c}_\text{4}$)} There exist $(N_{J},L_{J},E_{J},T_{J})$ satisfying the equations:
\begin{eqnarray}\label{eq2}
\left\{
\begin{split}
%N_{J}&=A-E_{J}CA-L_{J}^1C,\\
%T&_{J}=I-E_{J}C,\\
%L_{J}&=N_{J}E_{J},\\
%(E_{J}C-I)b_{J}&=\mathbf{0},\\
%L_{J}&=L_{J}^1+L_{J}^2,
N_{J}(I-E_{J}C)+L_{J}C+(E_{J}C-I)A=&0,\\
(T_{J}+E_{J}C-I)B=&0,\\
(E_{J}C-I)b_{J}=&0,
\end{split}
\right.
\end{eqnarray}
with detectable pair $(C,A-E_{J}CA)$, and Schur $N_{J}$.\\[1mm]
Then, if conditions ($\text{c}_\text{3}$) and ($\text{c}_\text{4}$) are satisfied, an UIO with the following structure:
\begin{eqnarray}\label{RUIO}
\left\{
\begin{split}
z_{J}^{+}=&N_{J}z_{J}+T_{J}Bu+L_{J}y,\\
\hat{x}_{J}=&z_{J}+E_{J}y,
\end{split}
\right.
\end{eqnarray}
exists for each $b_{J}$, $J\subset \left\lbrace 1,\hdots,p\right\rbrace $ with $\card(J)\leq 2q<p$, where $z_{J} \in\mathbb{R}^{n}$ is the observer state, $\hat{x}_{J} \in \mathbb{R}^{n}$ denotes the state estimate, and $(N_{J},L_{J},T_{J},E_{J})$ are the observer matrices satisfying \eqref{eq2}, see \cite{Ding2008} for further details. That is, system \eqref{RUIO} is an UIO observer for the system:
\begin{eqnarray}\label{s3}
\left\{
\begin{split}
x^{+}=&Ax+Bu+b_{J}a_{J},  \\
y=&Cx,
\end{split}
\right.
\end{eqnarray}
with unknown input $b_{J}a_{J}$ and known input $Bu$. It follows that the estimation error $e_{J}=\hat{x}_{J}-x$ satisfies the difference equation:
\begin{equation}\label{error_Dyn2}
e_{J}^{+}=N_{J}e_{J},
\end{equation}
with $N_{J}$ schur.
\begin{assumption}\label{asump1}
	There are at most $q$ attacked actuators, i.e.,
	\begin{equation}
	\card(W)\leq q < \frac{p}{2},
	\end{equation}
	where $q>0$ is the largest integer satisfying \emph{($\text{c}_\text{3}$)} and \emph{($\text{c}_\text{4}$)}.
\end{assumption}
\begin{lemma}\label{lm1}
Under Assumption \ref{asump1}, among all possible sets of $q$ actuators, at least one set includes all the attacked actuators.
\end{lemma}
\begin{lemma}\label{lm2}
Under Assumption \ref{asump1}, for each set of $q$ actuators, among all its supersets of $2q$ actuators, at least one superset includes all the attacked actuators.
\end{lemma}
\textbf{\emph{Proof:}} Lemma 1 and Lemma 2 follow trivially from the fact that $0<q<p/2$. \hfill $\blacksquare$\\[1mm]
Note that the existence of an UIO for each $b_{J}$ with $\card(J)\leq 2q$ implies that if $W \subseteq J$, the estimation error $e_{J} := \hat{x}_{J}-x$ satisfies
\begin{equation}
|e_{J}|\leq c_{J}\lambda_{J}^{k}|e_{J}(0)|,
\end{equation}
for some $c_{J}>0$, $\lambda_{J}\in(0,1)$, all $e_{J}(0)\in\mathbb{R}^{n}$ and all $k\geq 0$.\\[1mm]
Let Assumption \ref{asump1} be satisfied. We construct an UIO for each $J\subset\left\lbrace 1,\hdots,p\right\rbrace $ with $\card(J)=q$ and each set $S$ with $\card(S)=2q$. Then, by Lemma \ref{lm1}, there exists at least one set $\bar{J}\subset\left\lbrace 1,\ldots,p\right\rbrace $ with $\card(\bar{J})=q$ such that $W\subseteq\bar{J}$ and the estimate produced by the UIO for $\bar{J}$ is a correct state estimate. Thus, the estimates given by any $S \supset \bar{J}$ with $\card(S)=2q$ will be consistent with that given by $\bar{J}$. This motivates the following estimation strategy:\\[1mm]
For each set $J\subset\left\lbrace 1,\ldots,p\right\rbrace $ with $\card(J)=q$ and all $k\geq 0$, we define $\pi_{J}(k)$ as the largest deviation between $\hat{x}_{J}(k)$ and $\hat{x}_{S}(k)$ that is given by any set $S\supset J$ with $\card(S)=2q$:
\begin{equation} \label{54}
\pi_{J}(k):=\max_{S\supset J:\card(S)=2q}|\hat{x}_{J}(k)-\hat{x}_{S}(k)|,
\end{equation}
for all $k\geq 0$, and define the sequence $\sigma(k)$ as
\begin{equation}\label{55}
\sigma(k):=\underset{J\subset\left\lbrace 1,\hdots,p\right\rbrace :\card(J)=q}{\argmin} \pi_{J}(k).
\end{equation}
The estimate given by the set $\sigma(k)$ is a correct estimate, i.e.,
\begin{equation}\label{56}
\hat{x}(k):=\hat{x}_{\sigma(k)}(k),
\end{equation}
where $\hat{x}_{\sigma(k)}(k)$ denotes the estimate given by the set $\sigma(k)$, provides an exponential estimate of the system state. For simplicity and without generality, for all $J$ and $S$, $z_{J}(0)$ and $z_{S}(0)$ are chosen such that $\hat{x}_{J}(0)=\hat{x}_{S}(0)=\hat{x}(0)$. The following result summarizes the ideas presented above.
\begin{theorem}\label{th1}
Consider system \emph{(\ref{s1})}. Let conditions \emph{($\text{c}_\text{3}$)} and \emph{($\text{c}_\text{4}$)}, and  Assumption \ref{asump2} and Assumption \ref{asump1} be satisfied, and consider the multi-observer \emph{(\ref{54})-(\ref{56})}. Define the estimation error $e(k):=\hat{x}_{\sigma(k)}(k) - x(k)$; then, there exist constants $\bar{c}>0$ and $\bar{\lambda}\in(0,1)$ satisfying:
	\begin{eqnarray}
	|e(k)|\leq\bar{c}\bar{\lambda}^{k}|e(0)|\label{60},
	\end{eqnarray}
	for all $e(0)\in\mathbb{R}^{n}$, $k\geq 0$.
\end{theorem}
\textbf{\emph{Proof:}} By Lemma \ref{lm1}, there exists at least one set $\bar{J}$ with $\card(\bar{J})=q$ such that $\bar{J}\supset W$. By ($\text{c}_\text{3}$) and ($\text{c}_\text{4}$), for $J=\bar{J}\supset W$ with $\card(\bar{J})=q$, there exist $c_{\bar{J}}>0$ and $\lambda_{\bar{J}}\in(0,1) $, such that
\begin{equation}\label{63}
|e_{\bar{J}}(k)|\leq c_{\bar{J}}\lambda_{\bar{J}}^{k}|e(0)|,
\end{equation}
for all $e(0)\in\mathbb{R}^{n}$ and $k\geq0$. Moreover, for any set $S\supset\bar{J}$ with $\card(S)=2q$, we have $S\supset W$ $\forall k\geq0$; hence, by ($\text{c}_\text{3}$) and ($\text{c}_\text{4}$), there exist $c_{S}>0$ and $\lambda_{S}\in(0,1)$ such that
\begin{equation}\label{61}
|e_{S}(k)|\leq c_{S}\lambda_{S}^{k}|e(0)|,
\end{equation}
for all $e(0)\in\mathbb{R}^{n}$ and $k\geq 0$. Consider $\pi_{\bar{J}}$ in (\ref{54}). Combining the above results, we have that
\begin{eqnarray}
\begin{split}
\pi_{\bar{J}}(k)=&\underset{S\supset\bar{J}}{\max}|\hat{x}_{\bar{J}}(k)-\hat{x}_{S}(k)|\\
=&\underset{S\supset\bar{J}}{\max}|\hat{x}_{\bar{J}}(k)-x(k)+x(k)-\hat{x}_{S}(k)|\\
\leq&  |e_{\bar{J}}(k)|+\underset{S\supset\bar{J}}{\max}|e_{S}(k)|,
\end{split}
\end{eqnarray}
for all $k\geq 0$. From (\ref{63}) and (\ref{61}), we obtain
\begin{equation}\label{66}
\pi_{\bar{J}}(k)\leq 2c'_{\bar{J}}\lambda_{\bar{J}}^{'k}|e(0)|,
\end{equation}
for all $e(0)\in\mathbb{R}^{n}$ and $k\geq 0$, where \[c'_{\bar{J}}:=\underset{S\supset\bar{J}}{\max}\left\lbrace c_{\bar{J}}, c_{S}\right\rbrace,\lambda'_{\bar{J}}:=\underset{S\supset\bar{J}}{\max}\left\lbrace \lambda_{\bar{J}}, \lambda_{S}\right\rbrace.\] Note that $S\supset\bar{J}$ with $\card(S)=2q$, thus, from (\ref{55}), we have $\pi_{\sigma(k)}(k)\leq\pi_{\bar{J}}(k)$. From Lemma \ref{lm2}, we know that there exists at least one set $\bar{S}\supset\sigma(k)$ with $\card(\bar{S})=2q$ such that $\bar{S}\supset W$ $\forall k\geq 0$, and, by ($\text{c}_\text{3}$) and ($\text{c}_\text{4}$), there exist $c_{\bar{S}}>0$ and $\lambda_{\bar{S}}\in(0,1)$ such that
\begin{equation}\label{67}
|e_{\bar{S}}(k)|\leq c_{\bar{S}}\lambda_{\bar{S}}^{k}|e(0)|,
\end{equation}
for all $e(0)\in\mathbb{R}^{n}$ and $k\geq0$. From (\ref{54}), by construction \[
\begin{split}
\pi_{\sigma(k)}(k)=&\underset{S\supset \sigma(k):\card(S)=2q}{\max}|\hat{x}_{\sigma(k)}(k)-\hat{x}_{S}(k)|\\\geq&|\hat{x}_{\sigma(k)}(k)-\hat{x}_{\bar{S}}(k)|,
\end{split}
\] using the above lower bound on $\pi_{\sigma(k)}(k)$ and the triangle inequality, we have that
\begin{eqnarray}
\begin{split}
|e_{\sigma(k)}(k)|=&|\hat{x}_{\sigma(k)}(k)-x(k)|\\
=&|\hat{x}_{\sigma(k)}(k)-\hat{x}_{\bar{S}}(k)+\hat{x}_{\bar{S}}(k)-x(k)|\\
\leq&|\hat{x}_{\sigma(k)}(k)-\hat{x}_{\bar{S}}(k)|+|e_{\bar{S}}(k)|\\
\leq&\pi_{\sigma(k)}(k)+|e_{\bar{S}}(k)|\\
\leq&\pi_{\bar{J}}(k)+|e_{\bar{S}}(k)|,
\end{split}
\end{eqnarray}
for all $k\geq 0$. Hence, from (\ref{66}) and (\ref{67}), we have
\begin{eqnarray}\label{70}
|e_{\sigma(k)}(k)|\leq \bar{c}\bar{\lambda}^{k}|e(0)|,
\end{eqnarray}
for all $e(0)\in\mathbb{R}^{n}$ and $k\geq 0$, where $\bar{c}=3\max\left\lbrace c_{\bar{S}},c'_{\bar{J}}\right\rbrace $ and $\bar{\lambda}=\max\left\lbrace \lambda_{\bar{S}},\lambda'_{\bar{J}}\right\rbrace $. Inequality (\ref{70}) is of the form (\ref{60}), and the result follows. \hfill $\blacksquare$\\[1mm]
\textbf{Example 2} Consider the following system:
\begin{eqnarray}
\left\{ \begin{split}
x^{+}=&\left[ \begin{matrix}\label{e2}
0.5&0&0.1\\
0.2&0.7&0\\
1&0&0.3
\end{matrix}\right] x+\left[ \begin{matrix}
1&0&0\\
0&1&0\\
0&0&1
\end{matrix}\right] (u+a),\\
y=&\left[ \begin{matrix}
1&2&0\\2&1&3
\end{matrix}\right] x.
\end{split} \right.
\end{eqnarray}
It can be verified that a complete UIO for \eqref{e2} does not exist. However, partial UIOs exist for each $b_{J}$ with $\card(J)\leq 2$; then, $2q=2$, i.e., $q= 1$. We let $W=\left\lbrace 2\right\rbrace $, i.e., the second actuator is attacked. We let $u_{1},u_{2},u_{3}\in\mathcal{U}(-1,1)$, $a_{2}\in\mathcal{U}(-1,1)$, and $x_{1}(0),x_{2}(0),x_{3}(0)\in\mathcal{N}(0,1)$. We construct an UIO for each set $J\subset\left\lbrace 1,2,3\right\rbrace $ with $\card(J)=1$ and each $S\subset\left\lbrace 1,2,3\right\rbrace $ with $\card(S)=2$. Totally $\binom{3}{1}+\binom{3}{2}=6$ UIOs are designed and they are all initialized at $\hat{x}(0)=\left[ 0,0,0\right] ^{\top}$. The performance of the estimator is shown in Figure \ref{fig:1e}.

\section{Isolation of Actuator Attacks}
Once we have an estimate $\hat{x}(k)$ of $x(k)$, either using the complete observer in Section \ref{complete} or the partial multi-observer estimator in Section \ref{partial}, we can use these estimates, the system model \eqref{s1}, and the known inputs to exponentially reconstruct the attack signals. First, consider the complete observer in Section \ref{complete}. By construction, the estimation error $e = \hat{x} - x$ satisfies the difference equation \eqref{error_Dyn1} for some Schur matrix $N$. Note that $e = \hat{x} - x \Rightarrow x = \hat{x} - e \Rightarrow x^{+} = \hat{x}^{+} - e^{+}$. Then, the system dynamics \eqref{s1} can be written in terms of $e$ and $\hat{x}$ as follows:
\begin{eqnarray}\label{attack_estimation_1}
\left\{
\begin{split}
\hat{x}^{+} &= e^{+} + A(\hat{x} - e) + B(u+a),\\
&\hspace{25mm}\Downarrow\\
a&= B_{left}^{-1}(\hat{x}^{+}  -A\hat{x})-u + B_{left}^{-1}(e^+ - Ae), \\
\end{split}
\right.
\end{eqnarray}
because $B$ has full column rank, where $B_{left}^{-1}$ denotes the Moore-Penrose pseudoinverse of $B$. Therefore, because $e$ (and thus $e^+$ as well) vanishes exponentially, the following attack estimate:
\begin{eqnarray}\label{ru}
\hat{a}(k) &= B_{left}^{-1}(\hat{x}(k) - A\hat{x}(k-1))-u(k-1),
\end{eqnarray}
exponentially reconstructs the attack signals $a(k-1)$, i.e.,
\begin{equation}
	\begin{split}
\lim_{k\to\infty}\left( \hat{a}(k)-a(k-1)\right) =0.
	\end{split}
\end{equation}
Then, for sufficiently large $k$, we assume $\supp(a(k))=\supp(a(k-1))$, thus, the sparsity pattern of $\hat{a}(k)$ can be used to isolate actuator attacks at time $k$, i.e.,
\begin{equation}\label{su}
\hat{W}(k)=\supp(\hat{a}(k)),
\end{equation}
where $\hat{W}(k)$ denotes the set of isolated actuators at time $k$. Note that we can only estimate $a$ from $\hat{x}^{+}$ and $e^{+}$, which implies that we always have, at least, one-step delay. 

\begin{figure}[t]
	\includegraphics[width=0.5\textwidth]{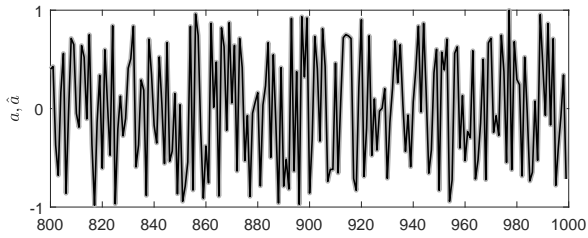}
	\caption{Estimated attack $\hat{a}$ converges to $a$ when $a\sim\mathcal{U}(-1,1)$. Legend: $\hat{a}$ (grey), $a$ (black)}
	\label{fig:e31}
	\centering
\end{figure}

\begin{figure}[t]
	\includegraphics[width=0.5\textwidth]{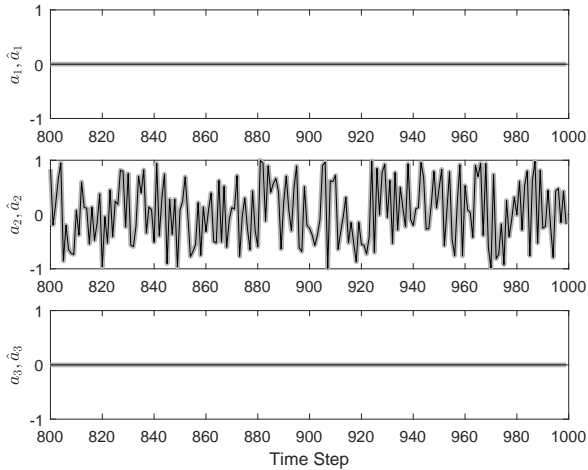}
	\caption{Estimated states $\hat{a}$ converges to $a$ when $a_{2}\sim\mathcal{U}(-1,1)$. Legend: $\hat{a}$ (grey), $a$ (black)}
	\label{fig:e35}
	\centering
\end{figure}

Next, consider the partial multi-observer estimator given in Section \ref{partial}. In this case, the attack vector $a$ can also be written as \eqref{attack_estimation_1} but the estimation error dynamics is now given by some \emph{nonlinear} difference equation characterized by the estimator structure in (\ref{54})-(\ref{56}). Let the estimation error dynamics be given by
\begin{equation}\label{c4}
e^{+}=f(e,x,a),
\end{equation}
for some nonlinear function $f:\mathbb{R}^n \times \mathbb{R}^n \times \mathbb{R}^p \rightarrow \mathbb{R}^n$. That is, the estimation error is given by some nonlinear function of the state and the attack signals. However, in Theorem \ref{th1}, we have proved that $e$ converges to the origin exponentially. Hence, the terms depending on $e$ and $e^+$ in the expression for $a$ in \eqref{attack_estimation_1} vanishes exponentially and therefore the attack estimate in \eqref{ru} exponentially reconstructs the attack signals. Again, the sparsity pattern of $\hat{a}(k)$ can be used to isolate actuator attacks using \eqref{su}.\\[1mm]
\textbf{Example 3}
Consider system (\ref{e1}) and the complete UIO in Example 1. Let $W=\left\lbrace 1\right\rbrace $,  $a\in\mathcal{U}(-1,1)$, $u\in\mathcal{U}(-1,1)$, and $x_{1}(0),x_{2}(0)\in\mathcal{N}(0,1)$. For $k\in\left[ 800,1000\right] $, we obtain the attack estimates $\hat{a}(k)$ using (\ref{ru}). The performance is shown in Figure \ref{fig:e31}.\\[1mm]
%Next, we let $W=\left\lbrace 1\right\rbrace $,  $a_{1}\in\mathcal{U}(-1,1)$. The performance is shown in Figure \ref{fig:e33}.\\[1mm]
%\begin{figure}[t]
%	\includegraphics[width=0.5\textwidth]{attackisonew-11.eps}
%	\caption{Estimated attack $\hat{a}$ converges to $a$ when $a_{1}\sim\mathcal{U}(-1,1)$. Legend: $\hat{a}$ (grey), $a$ (black)}
%	\label{fig:e33}
%	\centering
%\end{figure}
%\begin{figure}[t]
%	\includegraphics[width=0.5\textwidth]{attackisonew-1010.eps}
%	\caption{Estimated states $\hat{a}$ converges to $a$ when $a_{1}\sim\mathcal{U}(-10,10)$. Legend: $\hat{a}$ (grey), $a$ (black)}
%	\label{fig:e34}
%	\centering
%\end{figure}
\textbf{Example 4} Consider system (\ref{e2}) and the multi-observer estimator in Example 2. Let $q^{*}=1$, $W=\left\lbrace 2\right\rbrace $, $u_{1},u_{2},u_{3}\in\mathcal{U}(-1,1)$, $a_{2}\in\mathcal{U}(-1,1)$, and $x_{1}(0),x_{2}(0),x_{3}(0)\in\mathcal{N}(0,1^{2})$. For $k\in\left[ 800,1000\right] $, we obtain $\hat{x}(k)$ using (\ref{54})-(\ref{56}) and the vector $\hat{a}(k)$ from (\ref{ru}). The performance is shown in Figure \ref{fig:e35}. Note that $\hat{W}(k)=\{ 2 \}$, i.e., the second actuator is correctly isolated.

\section{Control}\label{control}
In this section, we propose a simple yet effective technique to stabilize the system by switching off the isolated actuators, i.e., by removing the columns of $B$ which correspond to the isolated actuators, and closing the loop with a multi-observer (observer) based output dynamic feedback controller. Indeed, we need the system to be stabilizable after switching off the isolated actuators. We first consider the case when a complete UIO exists (Section \ref{complete}), i.e., $\hat{x}$ is generated by \eqref{o}. We estimate $\hat{a}(k)$ using (\ref{ru}) and obtain $\hat{W}(k)$ from (\ref{su}). Again, let $B$ be partitioned as $B=\left[ b_{1},\hdots,b_{i},\hdots,b_{p}\right]$. Define $\bar{J}(k):=\left\lbrace 1,\ldots,p\right\rbrace \setminus\hat{W}(k)$ and $b_{\bar{J}(k)}$ as the matrix whose columns are $b_{i}$ for $i\in\bar{J}(k)$, i.e., $\bar{J}(k)\subset\left\lbrace 1,\ldots,p\right\rbrace $ is the set of isolated attack-free actuators and the columns of $b_{\bar{J}(k)}$ are the corresponding columns of $B$. Therefore, after switching off the set $\hat{W}(k)$ of actuators, the system has the following form:
\begin{equation}\label{d1}
	x^{+}=Ax+b_{\bar{J}(k)}\bar{u}
\end{equation}
where $\bar{u}\in\mathbb{R}^{\card(\bar{J}(k))}$ is the set of isolated attack-free inputs. Let $q^{\star}$ be the largest integer such that $(A,b_{J})$ is stabilizable for each set $J\subset\left\lbrace 1,\ldots,p\right\rbrace $ with $\card(J)\geq p-q^{\star}$ where $b_J$ denotes a matrix whose columns are $b_i$ for $i \in J$. We assume that at most $q^{\star}$ actuators are attacked. It follows that $p-q^{\star}\leq\card(\bar{J}(k))\leq p$. We assume the following.
\begin{assumption}\label{as3}
For any subset $J$ with cardinality $\card(J)=p-q^{\star}$, there exists a linear switching state feedback controller $\bar{u}=K_{\bar{J}(k)} x$ such that the closed-loop dynamics:
% $K_{J(k)}\in\left\lbrace K_{1},\ldots,K_{N}\right\rbrace=:\mathcal{K}$, $N=\binom{p}{p-q^{\star}}+\binom{p}{p-q^{\star}+1}+\ldots+\binom{p}{p}$ such that
\begin{equation}\label{c2}
x^{+}=(A+b_{\bar{J}(k)}K_{\bar{J}(k)})x,
\end{equation}
is GAS for $b_{\bar{J}(k)}$ arbitrarily switching among all $b_{J'}$ with $ J\subset J'\subset\left\lbrace 1,\ldots,p\right\rbrace$ and $p-q^{\star}\leq\card(J')\leq p$.
\end{assumption}
%\begin{remark}
%We do not give a method for designing the linear switching state feedback controller $\bar{u}=K_{\bar{J}(k)} x$. We refer the interested reader to, e.g., \cite{Daafouz2002} and references their, for design methods of linear switching controllers.
%\end{remark}
\begin{remark}
	We do not give a method for designing the linear switching state feedback controller $\bar{u}=K_{\bar{J}(k)} x$. We refer the interested reader to, e.g., \emph{\cite{Daafouz2002}} and references therein, for design methods of linear switching controllers.
\end{remark}
By switching off the set $\hat{W}(k)$ of actuators at time $k$, using the controller designed for the set $\bar{J}(k)$, and letting $\bar{u}=K_{\bar{J}(k)}\hat{x}$, the closed-loop system can be written as
\begin{eqnarray}\label{c1}
\begin{split}
x^{+} &= (A+b_{\bar{J}(k)}K_{\bar{J}(k)})x+b_{\bar{J}(k)}K_{\bar{J}(k)}e,\\
\end{split}
\end{eqnarray}
with estimation error $e=\hat{x}-x$ satisfying the difference equation (\ref{error_Dyn1}) for some Schur matrix $N$. Because $e(k)$ converges to zero exponentially, $e(k)$ in (\ref{c1}) is a vanishing perturbation. Hence, under Assumption \ref{as3}, it follows that $\lim_{k \rightarrow \infty}x(k) = 0$.

\begin{figure}[t]
	\includegraphics[width=0.5\textwidth]{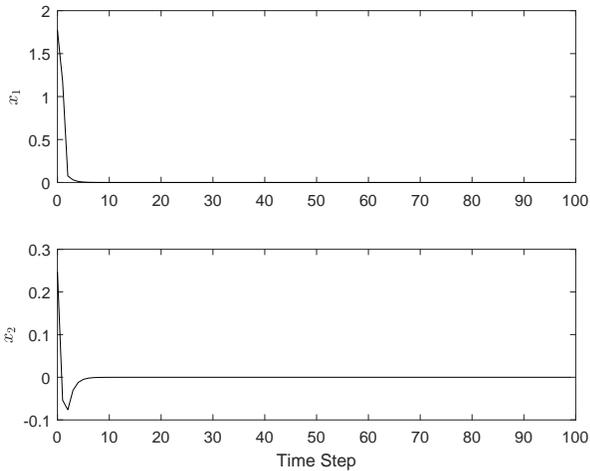}
	\caption{State trajectories when $a_{1}\sim\mathcal{U}(-1,1)$.}
	\label{fig:12a}
	\centering
\end{figure}

Next, assume that a complete UIO does not exist but partial UIOs exist for each $b_{J}$ with $\card(J)\leq 2q<p$ (Section \ref{partial}) and $q\leq q^{\star}$. We assume that at most $q$ actuators are attacked. We construct $\hat{x}(k)$ from (\ref{54})-(\ref{56}), estimate $\hat{a}(k)$ using (\ref{ru}), and obtain $\hat{W}(k)$ from (\ref{su}). After switching off the set $\hat{W}(k)$ of actuators, the system has the form (\ref{d1}) with $p-q\leq\card(\bar{J}(k))\leq p$. We assume the following.
\begin{assumption}\label{as4}
For any subset $J$ with cardinality $\card(J) = p-q$, there exists a linear switching state feedback controller $\bar{u}=K_{\bar{J}(k)} x$ such that the closed-loop dynamics:
% $K_{J(k)}\in\left\lbrace K_{1},\ldots,K_{N}\right\rbrace=:\mathcal{K}$, $N=\binom{p}{p-q^{\star}}+\binom{p}{p-q^{\star}+1}+\ldots+\binom{p}{p}$ such that
\begin{equation}\label{c3}
x^{+}=(A+b_{\bar{J}(k)}K_{\bar{J}(k)})x,
\end{equation}
is GAS for $b_{\bar{J}(k)}$ arbitrarily switching among all $b_{J'}$ with $ J\subset J'\subset\left\lbrace 1,\ldots,p\right\rbrace$ and $p-q \leq\card(J')\leq p$.
\end{assumption}
Then, by switching off the set $\hat{W}(k)$ of actuators at time $k$, using the controller designed for the set $\bar{J}(k)$, and letting $\bar{u}=K_{\bar{J}(k)}\hat{x}$, the closed-loop dynamics can be written in the form (\ref{c1}). Then, in this case, $e(k)$ is generated by some nonlinear difference equation of the form \eqref{c4}. Under Assumption \ref{as4}, the closed-loop dynamics (\ref{c1}) is Input-to-State Stable (ISS) with input $e(k)$ and some linear gain, see \cite{Sontag1997}. Moreover, in Theorem 1, we have proved that $e(k)$ converges to the origin exponentially uniformly in $x(k)$ and $a(k)$. The latter and ISS of the system dynamics imply that $\lim_{k \rightarrow \infty}x(k) = 0$ \cite{Jiang2001}.\\[1mm]
 \textbf{Example 5} Consider the following system:
\begin{eqnarray}
\left\{ \begin{split}
x^{+}=&\left[ \begin{matrix}\label{e2p}
1.5&0&0.1\\
0.2&0.7&0\\
1&0&0.3
\end{matrix}\right] x+\left[ \begin{matrix}
1&1&1\\
0&1&0\\
0&0&1
\end{matrix}\right](u+a),\\
y=&\left[ \begin{matrix}
1&2&0\\2&1&3
\end{matrix}\right] x,
\end{split} \right.
\end{eqnarray}
with $u=K_{\bar{J}(k)}\hat{x}$. Since $(A,b_{i})$ is stabilizable for $i\in\left\lbrace 1,2,3\right\rbrace $, we have $q^{\star}=2$. It can be verified that there does not exist a complete UIO for this system but partial UIOs exist for each $b_{J}$ with $\card(J)\leq 2$, then we have $q=1$. We let $W=\left\lbrace 1\right\rbrace $, and $a_{1}\in\mathcal{U}(-1,1)$. We construct $\binom{3}{1}+\binom{3}{2}=6$ UIOs and use the design method given in \cite{Daafouz2002} to build controllers for actuators $\left\lbrace 1,2\right\rbrace $, $\left\lbrace 1,3\right\rbrace $, $\left\lbrace 2,3\right\rbrace $, $\left\lbrace 1,2,3\right\rbrace $. Then, we use the multi-observer approach in Section \ref{partial} to estimate the state, reconstruct the attack signals, and control the system. The state of the system is shown in Figure \ref{fig:12a}.

\section{Conclusion}

We have addressed the problem of state estimation, attack isolation, and control for discrete-time LTI systems under (potentially unbounded) actuator false data injection attacks. Using a bank of Unknown Input Observers (UIOs), we have proposed an estimator that reconstructs the system state and the attack signals. We have proved that the designed estimator provides exponentially stable estimation errors for potentially unbounded attack signals. We used these estimates to control the system and isolate actuator attacks. We have provided simulations results to illustrate the performance of the results.


%\section*{Acknowledgment}



%\bibliographystyle{IEEEtran}
%\begin{thebibliography}{1}
%\end{thebibliography}

\bibliographystyle{IEEEtran}
\bibliography{Observer}
\end{document}